\documentclass{raa}
\usepackage{graphicx,times}
\usepackage{natbib}
\usepackage{aas_macros}
\usepackage{amssymb,amsmath}
\usepackage{xcolor}

\bibpunct{(}{)}{;}{a}{}{,}

\renewcommand{\vec}[1]{\boldsymbol{#1}}

\newcommand{\dif}{\mathrm{d}}

\def\kpc{{\rm kpc}}

\def\GeV{{\rm GeV}}

\def\km{{\rm km}}
\def\cm{{\rm cm}}
\def\sec{{\rm s}}
\def\yr{{\rm yr}}

\usepackage[a4paper=true,dvipdfm=true,pagebackref=true]{hyperref}
\hypersetup{pdftitle = The title of my PDF, pdfauthor = My name, pdfsubject= The subject, pdfkeywords = keyword1 keyword2 keyword3}
\hypersetup{colorlinks = true, linkcolor = green, anchorcolor = red, citecolor = blue, filecolor = red, pagecolor = red, urlcolor = red}

\begin{document}
\title{TeV cosmic-ray proton and helium spectra in the myriad model II}

\volnopage{ {\bf 2014} Vol.\ {\bf X} No. {\bf XX}, 000--000}
\setcounter{page}{1}

\author{Wei Liu\inst{1, 2}, Pierre Salati\inst{3}, Xuelei Chen\inst{1, 4}   }

\institute{$^1$ National Astronomical Observatories, Chinese Academy of Science, Beijing 100012, China \\
$^2$ University of Chinese Academy of Sciences, Beijing 100049, China {\it weiliu@bao.ac.cn }\\
$^3$  LAPTh, Universit\'e de Savoie, CNRS, BP 110, 74941 Annecy-le-Vieux,
France; {\it salati@lapp.in2p3.fr}\\
$^4$  Center of High Energy Physics, Peking University, Beijing 100871, China;
{\it xuelei@cosmology.bao.ac.cn}\\
\vs \no
{\small Received; accepted;\\Preprint number LAPTH-026/14}
}
\abstract{Recent observations show  that the cosmic ray nuclei spectra start to
harden above $\sim 10^2 ~\GeV$, in contradiction with
the conventional steady-state cosmic ray model. We had suggested 
that this anomaly is due to the propagation effect
of cosmic rays released from local young cosmic ray sources, the
total flux of the cosmic ray should be computed with the myriad model,
where contribution from sources in local catalogue is added to the background.
However, while the hardening could be elegantly explained in this model, 
the model parameters obtained from the fit skew toward a region with
fast diffusion and low supernova rate in the Galaxy, in tension
with other observations. In this paper, we further explore this model in
order to set up a concordant picture. 
Two possible improvements related to the cosmic ray
sources have been considered. Firstly, instead of the usual axisymmetric 
disk model, we considered a spiral model of source
distribution. Secondly, for the nearby and young sources 
which are paramount to the hardening, we allow for an energy-dependent escape time.
We find that major improvement comes from the energy-dependent escape 
time of the local sources, and with both modifications, not only the cosmic ray proton and
helium anomalies are solved, but also the parameters attain reasonable range values
compatible with other analysis.
}

\authorrunning{Wei Liu, Pierre Salati, Xuelei Chen}            
\titlerunning{TeV cosmic-ray proton and helium spectra in the myriad model II}
\maketitle

%
\section{Introduction} \label{intro}

It is generally accepted that most of the high energy
cosmic ray particles are accelerated in supernova shocks,
and form simple power-law spectra $q \propto {\cal R}^{-\alpha}$, 
where ${\cal R} \equiv p/Ze$
denotes the rigidity of particle
\citep{1977ICRC...11..132A, 1977DoSSR.234.1306K, 1978MNRAS.182..147B, 1978ApJ...221L..29B}.
After being injected into the interstellar space, the particles are frequently scattered
by magnetic irregularities in the Galaxy. This could be described
approximately as a diffusion process with spatial
diffusion coefficient $K \propto {\cal R}^{\delta}$
\citep{1990acr..book.....B, 2002astro.ph.12111M}.
The diffusion volume is called the magnetic halo. Once the cosmic rays
reach a steady state in the halo, the observed cosmic-ray fluxes are
given in a large energy range by $\Phi \propto q/K$, i.e. the
spectrum scales with energy as a single power law $E^{-(\alpha +\delta)}$,
where both $\alpha$ and $\delta$ are constant.

Recently the observations of the CREAM\citep{2010ApJ...714L..89A, 2011ApJ...728..122Y} and
PAMELA\citep{2011Sci...332...69A} experiments indicated a hardening of the cosmic nuclei
spectra above 250 GeV/nucleon, which extends to TeV energies, though the low energy spectrum
measured by the AMS-02 experiment \citep{ams2, 2014arXiv1402.0467C} differs somewhat from
the PAMELA result, and for the AMS-02 result the hardening is not as significant as the
PAMELA one.  This problem will be further checked by other experiments,
e.g. the AS$\gamma$ experiment \citep{2011AdSpR..47..629T, 2011ASTRA...7...15A}
in Yangbajing, Tibet.

A number of different models have been proposed to explain the PAMELA and CREAM
anomaly. Most of these focus on the change of energy dependence of
either the injection spectrum
$q(E)$\citep{2012PhRvL.108h1104M,2011ApJ...729L..13O,2010ApJ...725..184B},
or the diffusion coefficient $K(E)$\citep{2009ApJ...697..106A,2012PhRvL.109f1101B}.
A local variation
in K proposed by \citet{2012ApJ...752L..13T} could induce similar effect.
Besides, \citet{2012JCAP...01..010B} invokes an unusual strong spallation of the
species on the Galactic gas, but this has been criticized by \citet{2012ApJ...752...68V}.
More recently \citet{2013arXiv1308.1357T} considered
the diffusive re-acceleration effect, through which
the injected energy spectrum can be modified at low energies.

\citet{2013A&A...555A..48B} proposed that the excess of
cosmic-ray nuclei at high energy comes from particular configuration
of local sources. To compute the flux of the primary cosmic
ray nuclei, instead of the regular steady-state transport model, the
myriad-source model\citep{2003ApJ...582..330H} is employed . In this model
the cosmic ray flux is computed in a time-dependent transport framework.
For the young and nearby sources,
a detailed and complete catalogue was constructed using  current survey data, and its
contribution was calculated separately.  New values of the
transport parameters were obtained which fit the proton and helium excess
as well as the B/C ratio.

However, although this model could reproduce the observed spectra,
it is not completely satisfactory. The best fit model parameters  indicates
either a thin magnetic halo or a low supernova rate in the Galaxy, but these
are in tension with other observations, for example the
$\gamma-$ray and synchrotron emission
\citep{2010ApJ...722L..58S, 2012JCAP...01..049B, 2013JCAP...03..036D}. All these
results favor a medium size magnetic halo, but for
such a halo, e.g. $L \sim 4 \; \kpc$, a low supernova explosion rate is required,
at nearly the lower limit of the observations\citep{2006Natur.439...45D}.
It might be that the contribution from the background sources was overestimated. Another
possible problem is that the process for the cosmic ray particles escaping from
local supernova remnants (SNRs) was incorrectly modelled.

In this paper we consider two possible improvements to this model. In previous
studies \citep{2013A&A...555A..48B}, we have assumed that the background sources are distributed
axisymmetrically in the Galactic disk.
In fact, the sources, i.e. supernova remnants are likely distributed along the
spiral arms, and to our knowledge the solar system is located between two of them. This 
difference in the source distribution may affect the result.
Moreover, we shall also take the macroscopic size of the local sources and
the energy-dependent escape time into
account\citep{2012MNRAS.421.1209T, 2013arXiv1304.1400T}.
We shall investigate if the fitting to cosmic ray data can be improved, 
esp. with a higher supernovae explosion rate. 
We also incorporate the more recent and more precise 
AMS-02 data\citep{ams2, 2014arXiv1402.0467C} in our analysis.  

The paper is arranged as follows: section \ref{model} and \ref{impro}
introduce cosmic-ray propagation model and sources respectively.
The results are shown in section \ref{res}. Finally we give our discussions and
conclusions in section \ref{con}.

\section{Cosmic Ray Propagation Model} \label{model}
The galactic supernova explosions which inject cosmic ray particles into the interstellar space
can be regarded as a stationary random process. During the average lifetime of the cosmic ray
particles the number of supernova explosions is large, as a result a nearly steady
average flux is established. However, for short time scales (still large compared with
human history) and on scales relevant for observation (solar system),
statistical fluctuations can still be large
\citep{2012A&A...544A..92B}, and the contribution of young and nearby sources
could result in significant deviation from the average. This could be a possible reason for the
observed excess at high energies. To address this possibility,
we separate the cosmic ray sources into two components: the remote or aged
SNRs which produce a nearly steady background,
and the recent and nearby sources
which could produce a local deviation in the spectrum. We model the local
population by using the information collected from current SNR surveys.
Similar ideas were also explored
by \citet{Erlykin2012371} and \citet{2012MNRAS.421.1209T}, we
expanded it to the whole energy range from tens of GeVs up to a few
PeV  \citep{2013A&A...555A..48B}.
The key idea here is that the nearby sources is used to explain the
spectral hardening at high energies, whereas the bulk of the remote and old
sources account for the fluxes below 250 GeV/nuc. The more energetic particles
spend less time in the magnetic halo, making the
contribution from local and recent SNRs more important at high energies.

We are interested mainly in the high energy distribution,
so diffusive re-acceleration is neglected, as it acts mostly on low energy particles.
If we define $\psi \equiv \dif n/\dif T$ as the number density per unit volume and unit
kinetic energy of a given cosmic ray species, the diffusion equation is
\begin{equation}
\frac{\partial \psi}{\partial t} +\partial_z(V_c \psi) -K(E)\Delta \psi
+ 2h\delta(z)\ \Gamma_{\text{sp}} \psi= Q_{\text{acc}},
\label{CRP_eq}
\end{equation}
where $V_c$ is the convective velocity at which cosmic ray particles are blown
away from galactic disc by stellar wind. The spatial diffusion coefficient is
$K(E) = \kappa_0 \; \beta \; {\cal R}^\delta$, where $\kappa_0$ is a normalization constant
and $\beta = {v}/{c}$ denotes the particle velocity $v$ in units of speed of light $c$.
Generally the diffusion coefficients form a tensor and are
position-dependent, but here for simplicity we assume diffusion to be
uniform and isotropic within the magnetic halo.
The total cosmic ray flux is $\Phi = {v \psi}/(4 \pi) $. The magnetic
halo inside which cosmic
ray particles diffuse is assumed to be a flat cylinder,
whose thickness $2L$ is an unknown parameter to be determined from fitting to the data.
A uniform Galactic disk is located in the middle of halo and
its thickness is about $200$ pc. The radius of the
magnetic halo is usually set to be equal to Galactic radius $R = 20$ kpc.
Beyond the magnetic halo, the magnetic field drops rapidly,
and the cosmic ray particles are no longer
confined. Customarily, at the boundary the freely escape condition is assumed. The last
term in the left-hand side of Eq. \ref{CRP_eq} is the spallation term, with
the collision rate $\Gamma$  given by
\begin{equation}
\Gamma_{\text{sp}} = v(\sigma_{\text{pH}} n_{\text{H}} + \sigma_{\text{pHe}} n_{\text{He}}),
\end{equation}
in the case of cosmic ray protons.
The average densities of hydrogen and helium $n_{\text{H}}$ and $n_{\text{He}}$ in the disk are
$0.9$ and $0.1 \, \cm^{-3}$ respectively. The cross section $\sigma_{\text{pH}}$ is given in
\citet{2010JPhG...37g5021N}, and $\sigma_{\text{pHe}}$ is assumed to be
$4^{2.2/3} \sigma_{\text{pH}}$\citep{Norbury2007187}.
For helium cross sections the same scaling factor is adopted.

The sources are assumed to be point-like:
\begin{equation}
Q_{\text{acc}}(\vec{x}_S, t_S) = \sum_{i \in \mathcal{P} } q_i\ \delta^3 (\vec{x}_S -\vec{x}_i) \delta(t_S -t_i).
\end{equation}
where $q_i, \vec{x}_i, t_i$ denote the injection amount, position and time for the $i$th explosion.
For simplicity, we assume $q_i$ to be identical for all the sources and given by
\begin{equation} \label{inj_s}
q_i^j(p) = q^0_j \left(\frac{p}{1 \text{GeV}/\text{nuc}} \right)^{-\alpha_j},
\end{equation}
for the $j$th element. The parameters $q^0_j$ and $\alpha_j$
for proton and helium are determined from parameter fitting.
The solution of the transport equation~(\ref{CRP_eq}) can be written in
terms of Green function as
\begin{equation}
\psi(\vec{x}, t) = \int_{-\infty}^t \dif t_S \int_{MH} \dif^3 \vec{x}_S \ \mathcal{G}_p(\vec{x}, t\leftarrow \vec{x}_S, t_S)\ Q_{acc}(\vec{x}_{S}, t_{S}).
\end{equation}

The diffusion equation can be solved by numerical integration using the
 GALPROP\citep{1998ApJ...509..212S, 2007ARNPS..57..285S} code and
its succeeding DRAGON\citep{2008JCAP...10..018E, 2010APh....34..274D} package.
Here we consider another approach, the semi-analytical
model\citep{2001ApJ...555..585M, 2002A&A...394.1039M, 2002astro.ph.12111M}.
As the diffusion equation is linear, we can write the cosmic ray flux  as
\begin{equation}
\Phi = \Phi_{\text{cat}} + \Phi_{\text{ext}},
\end{equation}
where $\Phi_{\text{cat}}$ is the contribution of the nearby and recent sources given in the
catalogue, and $\Phi_{\text{ext}}$ is the flux from the background sources,
which can be approximated
by its time average $\bar{\Phi}_{\text{ext}}$ \citep{2012A&A...544A..92B}.
The nearby recent supernova explosions can
produce fluctuations in cosmic ray fluxes. With data from
current surveys, a catalogue of nearby SNRs can be constructed. However, SNRs have limited
lifetimes, and faint sources may still be missed.
Pulsars are usually regarded as relics of
supernova explosions, and are good tracers of aged SNRs.
They can be added to known SNRs as a complementary catalogue.
A list of cosmic ray sources derived from the
Green catalogue\citep{2009BASI...37...45G} and ATNF pulsar database\citep{2005AJ....129.1993M}
was presented in \citet{2013A&A...555A..48B}, with the millisecond pulsars
associated with known SNRs removed to avoid double counting. In this list, the radial
distance extends up to $2$ kpc from the Sun, and the
upper limit on the age is set to $30, 000$ years,
within which there are $30$ sources in total. This is consistent
with a supernova explosion rate of
3 per century\citep{2012A&A...544A..92B} which could be a local fluctuation
 on the high side since the Earth
is located between two nearby spiral arms.

\section{Improvements of the Source Model} \label{impro}
\subsection{Spiral distribution of CR sources}
Previously the cosmic ray source distribution was usually
assumed to be azimuth-symmetric in Galaxy. This is appropriate
when the diffusion distance is much larger than the characteristic
scale between spiral arms. But when evaluating cosmic ray fluxes
at high energies where local sources could play a dominant role,
the specific position of the solar system and its local environment may be important.

\begin{figure}
\centering
\includegraphics[width=0.6\textwidth]{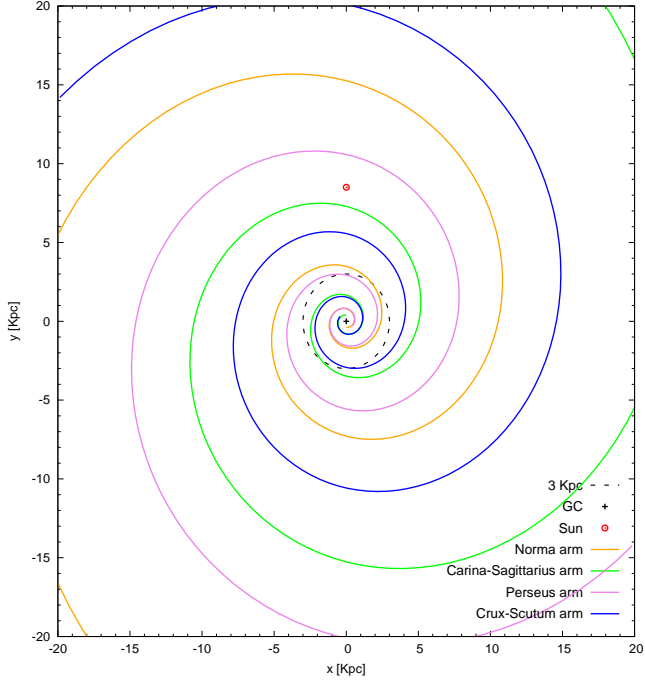}
\caption{
The Galaxy is assumed to have four spiral arms, with the Sun lying between 
the Carina-Sagittarius and Perseus arms, about $8.5$ kpc away from the Galactic 
center \citep{2006ApJ...643..332F}.
}
\label{fig:spiral_arm}
\end{figure}
The Milky Way Galaxy is a typical spiral galaxy, and the
high density gas inside spiral arms triggers rapid star formation, so the cosmic ray
sources are also highly correlated with the spiral arms. The spiral distribution
of sources 
have been considered in a some recent studies\citep{2003NewA....8...39S,
2002PhRvL..89e1102S, 2009PhRvL.103k1302S, 2012arXiv1210.1423E, 2012JCAP...01..010B}.
There are still some uncertainties in the structure and pattern speed of the spiral arms,
owing to our position in the Galaxy. While the outer part of the Milky Way seems to have
four arms, for the inner part the number of arms is still being debated. 
The different measurements for the spiral structure
and number of spiral arms are reviewed in \citet{1995ApJ454119V, 2002ApJ566261V}
and \citet{1998ggs..book.....E}.
In this paper, we adopt the spiral model given in \citet{2006ApJ...643..332F}.
Similar models were also adopted by \citet{2012JCAP...01..010B} and \citet{2013PhRvL.111b1102G}.
In this model, the whole Galaxy is assumed to be made of four major arms spiralling outward from the
Galactic Center, as featured in Fig. \ref{fig:spiral_arm}. The loci of the ith arm is on a
logarithmic spiral defined by the relation $\theta(r) = k^i\ln(r/r_0^i) +\theta_0^i$,
where the parameters $k^i$, $r_0^i$ and $\theta_0^i$ are borrowed from \cite{2006ApJ...643..332F}.
For the radial distribution of sources, the parameterizations
of \citet{2004A&A...422..545Y} is applied,
\begin{equation}
\rho(r) = \left(\frac{r +R_1}{R_{\odot} +R_1}\right)^{a}
\exp\left[-b\left( \frac{r -R_{\odot}}{R_{\odot} +R_1}\right) \right],
\end{equation}
with the best-fit values $a = 1.64, b = 4.01$ and $R_1 = 0.55$ kpc. The Solar System
is located between the Carina-Sagittarius and Perseus spiral arms, with
$\vec{R}_{\odot} = \{0 , 8.5 , 0\}$ in units of kpc. Along the spiral arm,
there is a spread in radial coordinate with normal distribution
\begin{equation}
f_{i} = \frac{1}{\sqrt{2\pi}\sigma}\exp \left( -\frac{(r-r_{i})^2}{2\sigma^2} \right)
\qquad  \text{i} \in [1,2,3,4]
\end{equation}
where $r_{i}$ is the inverse function of loci of the i-th spiral arm and
standard deviation $\sigma$ is taken to be $0.07 \; r_{i}$. The vertical distribution from
Galactic plane is a decreasing exponential function with mean $z_0 = 100$ pc.
Different measurements have been used to determine the pattern speed. Here we
take $\Omega_{\odot} -\Omega_{p, max}$ to be
$7.7 \; \km \, \sec^{-1} \, \kpc^{-1}$\citep{2003NewA....8...39S}.
\citet{2006ApJ...643..332F} also modified their model in the inner 3 kpc 
of the Galaxy to account for the more axisymmtric core,
 but for our calculation at the solar position
this makes little difference, so we shall simply use the 
spirals for calculation.

\subsection{Energy-dependent escape}
In the usual treatment, the inherent size and duration of the source are neglected, all the cosmic
ray particles are assumed to be released into the interstellar space
instantly once a supernova explodes. This
approximation is reasonable when we are considering scales much larger than the
size and duration of sources. However, as we noted above, at high energies the role of
recent nearby sources may be important, and the observed cosmic ray anomaly may be a local
fluctuation, then these neglected factors may also need to be taken into account
\citep{2012MNRAS.421.1209T, 2013arXiv1304.1400T}.

According to the diffusive shock acceleration theory, charged particles are accelerated
during their repeated crossings of the shock front. They are confined within supernova shock
by magnetic turbulence until their upstream diffusion length $l_d$ is larger than the
shock radius, which is growing with time.
The diffusion length is given by $l_d = D_s/u_s$, where $D_s$ is the diffusion coefficient in the
upstream region and $u_s$ is the shock velocity. In the Bohm limit, the diffusion
coefficient $D_s$
in the upstream region increases linearly with energy $E$, $D_s(E) \propto E$. Thus in general
the most energetic particles escape from the acceleration region earlier.
However the details of how the cosmic ray escape from the source are still not well understood.

The escape time is generally parameterized as
\begin{equation}
t_{esc}({\cal R}) = t_{sed} \left(\frac{{\cal R}}{{\cal R}_{max}} \right)^{-1/\gamma},
\label{eq:gamma}
\end{equation}
where $t_{sed} =  500$ years is the onset time of the Sedov phase, and the maximum
rigidity ${\cal R}_{max} = 1$ PV. The escape index $\gamma$ is a positive constant and
determines the span of the escape time. When $\gamma \gg 1$, the escape is very close to
instantaneous injection into space. Usually this parameter lies between $1$ and $3$. When the
shock is too weak to accelerate particles and the turbulence level in the upstream region can no
longer hold them, the rest of the cosmic ray particles are released all at once.
This is assumed to happen at $10^5$ years, so the cosmic ray escape time is taken
to be
\begin{equation}
T_{esc}(E) = \text{min}(t_{esc}, 10^5 \yr).
\end{equation}
Along with shock expansion, the shock radius increases according to the Sedov relation
\begin{equation}
R_{esc}(E) = 2.5 \, u_0 \, t_{sed} \, \left\{ \left(\frac{T_{esc}}{t_{sed}} \right)^{0.4}  -0.6 \right\},
\end{equation}
where $u_0 = 10^7$ m$/$s represents the initial velocity of the shock at time $t_{sed}$.
If cosmic rays are assumed to escape from the surface, which is supposed to be spherically symmetric,
the source term  in equation~(\ref{CRP_eq}) turns out to be
\begin{equation}
Q(E,t,r) = \frac{q(E)}{A_{esc}} \, \delta(t -T_{esc}) \, \delta(r -R_{esc}),
\end{equation}
where $A_{esc} = 4\pi R^2_{esc}$ is the surface area of the SNR at the escape time $T_{esc}$,
and $r$ denotes the distance to the SNR center.

\section{Results} \label{res}

We now try to find the best-fit parameter values by comparing the model prediction on
proton and helium spectra with the data from the
AMS-02\citep{ams2} and CREAM\citep{2010ApJ...714L..89A} experiments. We perform the fit
in the energy range from $50$ GeV/nuc to $100$ TeV/nuc, where the solar
modulation can be safely neglected. The quality of the fit to
the data is analyzed quantitatively by $\chi^2/\mathrm{d.o.f.}$, with the
proton and helium data,
$\chi^2=\chi^2_{\text{p}} +\chi^2_{\text{He}}.$ Our cosmic ray propagation
model is defined by the transport parameters $K_0$, $\delta$, $V_c$ and $L$.
All of these are restricted within a range that is consistent with the
secondary-to-primary B$/$C measurements\citep{2001ApJ...555..585M}. In addition, we have
parameters which specify the source properties, including
$q^0_{\text{p}}$, $q^0_{\text{He}}$, $\alpha_{\text{p}}$,
$\alpha_{\text{He}}$ in Eq.(\ref{inj_s}), and the average supernova explosion rate $\nu$.
Finally, for the energy-dependent escape model, one more parameter, the
escape index $\gamma$, also needs to be included. The parameters and goodness of fit
for the models considered in this paper is listed in Table.\ref{ta:model}.

\begin{table*}
\caption{The sets of cosmic ray injection and propagation parameters and the goodness of fit
for different models considered in this paper. 
The Aa (axisymetric) and A(spiral) models assume point 
sources and instantaneous injection, while the B,C,D and E models include effects of 
finite size and energy-dependent escape. The index $\gamma$ specifies 
energy-dependence of escape time (see Eq.(\ref{eq:gamma})).
}
\label{ta:model}
\centering
\begin{tabular}{|c|c|c|c|c|c|c|}
\hline\hline
Model & A & Aa & B & C & D & E \\
\hline
Diffusion coefficient normalization $K_0$ & 0.0112 & 0.0112 & 0.0112 & 0.0112 & 0.0112 & 0.0112 \\
\hline
Diffusion spectral index $\delta$ & 0.7 & 0.7 & 0.7 & 0.7 & 0.7 & 0.7 \\
\hline
Magnetic halo half thickness L [kpc] & 4 & 4 & 4 & 4 & 4 & 4 \\
\hline
Convective Velocity $V_c$ [km s$^{-1}$] & 12 & 12 & 12 & 12 & 12 & 12 \\
\hline
SN explosion rate $\nu$ [century$^{-1}$] & 0.8 & 0.8 & 1.2 & 1.4 & 1.7 & 1.5 \\
\hline
SN proton injection number $q^0_{\text{p}}$  [$10^{52}$ GeV$^{-1}$]  & 2.527 & 1.773 & 
1.355 & 1.188 & 0.933 & 1.068 \\
\hline
SN proton spectral index $\alpha_{\text{p}}$  & 2.2 & 2.17 & 2.154 & 2.159 & 2.150 & 2.151 \\
\hline
SN helium injection number $q^0_{\text{He}}$ [$10^{51}$ GeV$^{-1}$]  & 1.475 & 1.044 & 0.768 & 
0.734 & 0.554 & 0.648 \\
\hline
SN helium spectral index $\alpha_{\text{He}}$   & 2.07 & 2.04 & 2.012 & 2.037 & 2.017 & 2.024 \\
\hline
Energy-dependence index $\gamma$  & $\infty$ & $\infty$ & 2 & 3 & 2.36 & 2.37 \\
\hline
Goodness of Fit $\chi^2/$dof  & 2.19 & 1.93 & 1.708 & 1.552 & 1.751 & 1.469 \\
\hline\hline
\end{tabular}
\medskip
\end{table*}

In our previous work\citep{2013A&A...555A..48B}, 
we studied several configurations, but 
the supernova explosion rates were close to the lower observational limit,
unless $L$ is extremely small. 
In this paper, we fix the propagation parameters
to the MED configuration of \citet{2004PhRvD..69f3501D,2013A&A...555A..48B}, where the
vertical boundary of halo $L$ has the plausible value of 4 kpc, and also best fits the
B/C data\citep{2001ApJ...555..585M}. We perform a scan over supernova explosion
rate, which is required to be larger than $0.8$ century$^{-1}$. Once the
transport parameters and the explosion rate are fixed, source parameters
$q^0_{\text{p}}$, $q^0_{\text{He}}$, $\alpha_{\text{p}}$ and $\alpha_{\text{He}}$ are automatically
adjusted to find the best-fit for both proton and helium data.

\begin{figure}
\centering
\includegraphics[width=0.6\textwidth]{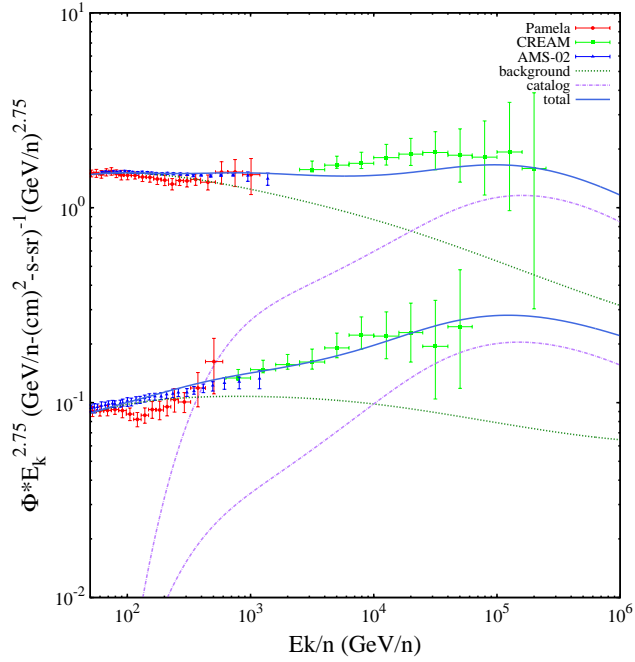}
\caption{
\label{fig:fitA}
The best-fit model A(see table \ref{ta:model}) to the AMS-02 and
CREAM data under spiral distribution of background sources, where
local sources are assumed to be point-like with instantaneous injection. The
propagation parameters are configured to the MED case, where $L$ is $4$ kpc, and
explosion rate of $0.8$ per century is assumed. Proton(upper curve) and helium(lower curve)
spectra are featured in the energy range extending from $50$ GeV/nuc to $100$ TeV/nuc.  
The contribution from the background sources corresponds to the green dotted lines. The
blue solid lines show the total flux while the purple dash-dotted curves indicate the flux
from the sources of the catalog.
}
\end{figure}

First, we study the spiral distribution model with  instantaneous injection
for local sources (model A). The average supernova explosion rate is taken as $0.8$ per
century. The results are presented in Fig.~\ref{fig:fitA}, where the upper
curves are for proton, and lower curves for helium.
The background sources contribution $\Phi_{\text{ext}}$
is plotted as the green dotted lines, the
nearby young sources contribution $\Phi_{\text{cat}}$ as the purple
dash-dotted lines, and the total flux as the blue solid lines. 
The background flux $\Phi_{\text{ext}}$ is dominant below $\sim$ 100 GeV, 
where the AMS-02 data dominate the fit. As energy increases,
the background contribution shrinks, while the contribution from nearby young sources raises. 

\begin{figure}
\centering
\includegraphics[height=0.6\textwidth]{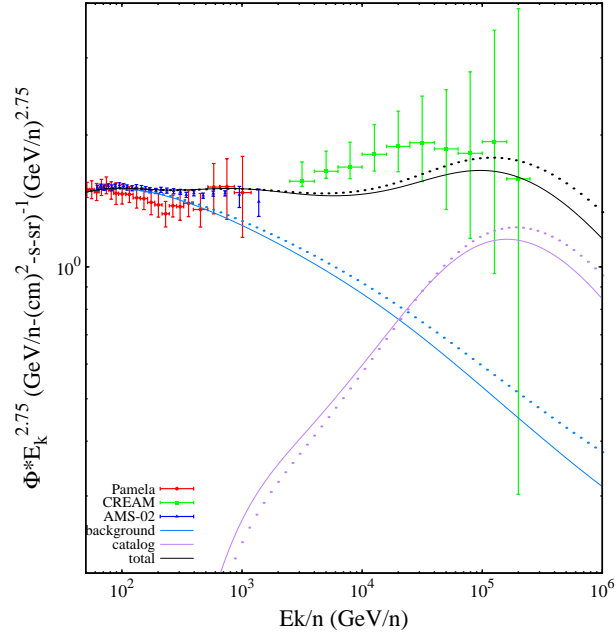}
\caption{
\label{fig:flux_comp}
Various contributions to the proton flux have been calculated in the spiral and axisymmetric
source models for comparison purposes. The solid and dotted curves correspond
to the spiral and axisymmetric configurations A and Aa of 
Table~\ref{ta:model} respectively. The black lines feature
the total flux which is a sum of the contributions from the background sources (blue) and from the
catalog of local objects (purple).
}
\end{figure}

As shown in Table.\ref{ta:model}, 
we find that compared with the original axisymmetric model  \citep{2013A&A...555A..48B},
 the goodness of fit for the spiral model is not improved. To understand this,
in Fig.~\ref{fig:flux_comp} we compare the proton flux for
models A (spiral) with model Aa(axi-symmetric).
Both models are based on the same cosmic ray propagation parameters,
and have the same average supernova explosion rate $\nu$. The solid curves
correspond to the spiral model A whereas the dotted curves feature the 
axi-symmetric case model Aa. For the low energy part, the two models have almost the same
background contribution. 
Now the Earth is located between the Carina-Sagittarius and Perseus arms, the number
of sources which contribute to $\Phi_{\text{ext}}$ are less numerous in the spiral model
than for the axisymmetric model. The latter tends to overpopulate the void
inside which the solar system is located and leads
to a slightly larger background flux above $\sim$~1~TeV.
This effect is counterbalanced by a smaller contribution $\Phi_{\text{cat}}$ from the local sources
up to an energy of 10~TeV, so the best-fit value for $q^0_{\text{p}}$ is larger in model
A than in model Aa. Conversely, the high-energy data points demands
a slightly softer injection spectrum and a larger value for the
index $\alpha_{\text{p}}$ in model A than
Aa. However, the difference between model A and model Aa is not very significant. Perhaps
more important is the behavior of the background flux $\Phi_{\text{ext}}$ at high energy 
in the presence of spiral arms, as the energy increases, it drops faster than the simple power-law as 
was assumed in \citet{2012MNRAS.421.1209T, 2013arXiv1304.1400T}.

\begin{figure}
\centering
\includegraphics[height=0.6\textwidth]{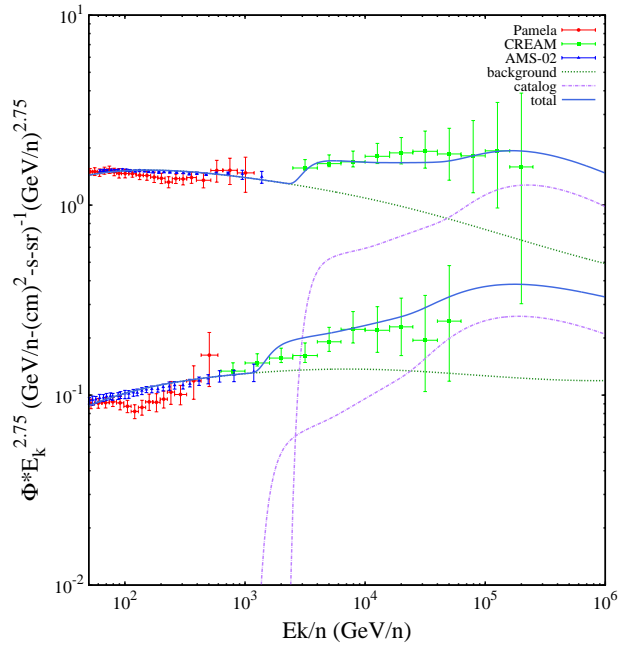}
\caption{
Model B assumes energy-dependent escape of cosmic rays from local sources.
The escape index $\gamma$ is set equal to $2$ as default value while the explosion
rate is equal to $1.2$ per century. The cosmic ray propagation parameters are
configured to the MED model.
}
\label{fig:fitB}
\end{figure}

\begin{figure}
\centering
\includegraphics[width=0.6\textwidth]{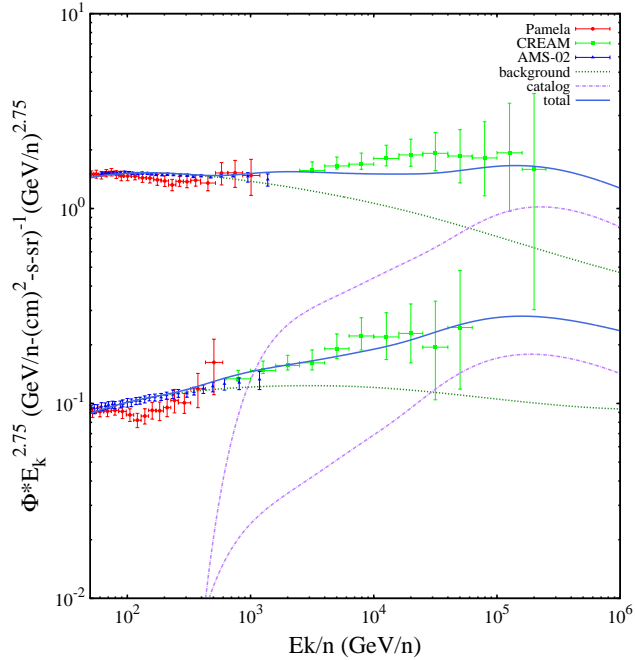}
\caption{
Another best-fit model C with same constraints as before.
The escape index $\gamma$ is set equal to $3$ as default value and the explosion
rate is equal to $1.4$ per century.
}
\label{fig:fitC}
\end{figure}

Next we consider the energy-dependent injection from local sources,
the background sources are assumed to be spirally distributed. As we discussed earlier,
modifying the escape index $\gamma$ changes the energy range where the local component 
comes into play. When $\gamma$ gets large,
energy-dependent escape gradually approaches an instantaneous injection taking place at
the start of the Sedov phase, and sources then behave as point-like objects
(the radius is only $\sim$ 5~pc).
As $\gamma$ decreases, the span of escape time becomes longer. The most
energetic particles escape first from the SNR. When the injection timescale becomes
comparable with the upper limit of 30,000 years on the age of the local sources, only the
most energetic cosmic rays are released and make it to Earth.
We explored different values of $\gamma$, and found at $\gamma$ values at  $2 \sim 3$ 
good fits can be obtained. In the best-fit models B and C, the MED propagation
parameters are assumed, with $\gamma=2$ and $\gamma=3$ for model B and C, respectively.
We explore here how the local flux $\Phi_{\text{cat}}$
reacts to a change in the escape index $\gamma$. The results are shown in
Figs.~\ref{fig:fitB} and \ref{fig:fitC}. There is a low energy cut-off  
in the local contributions to the proton and helium fluxes
in the case of model B where $\gamma = 2$ has been assumed. This is due to the 
energy-dependent cosmic ray release mechanism we discussed earlier: the lower energy 
cosmic ray particles from young sources have not yet been released or reached Earth.
As a result of this low energy cut-off, there is a also a kink at this energy 
in the spectrum predicted by the model. In model C, with $\gamma=3$, 
the cutoff is not as sharp and the local flux starts to contribute from lower energies.
The kink in total flux at low energy in model B (Fig.~\ref{fig:fitB})
disappeared in model C (Fig.~\ref{fig:fitC}). 

We find that when taking the energy-dependent escape into account, the goodness of fit
is markedly improved. 
For completeness, we have also varied the supernova explosion rate between models B
and C. A larger value of $\nu$ translates into a larger abundance of sources contributing
to the overall signal. The amount of cosmic rays injected by a single object decreases
as is clear shown in Table~\ref{ta:model}. Although the effect is marginal, the contribution from
local sources drops gradually with increasing explosion rate.

\begin{figure}
\centering
\includegraphics[width=0.6\textwidth]{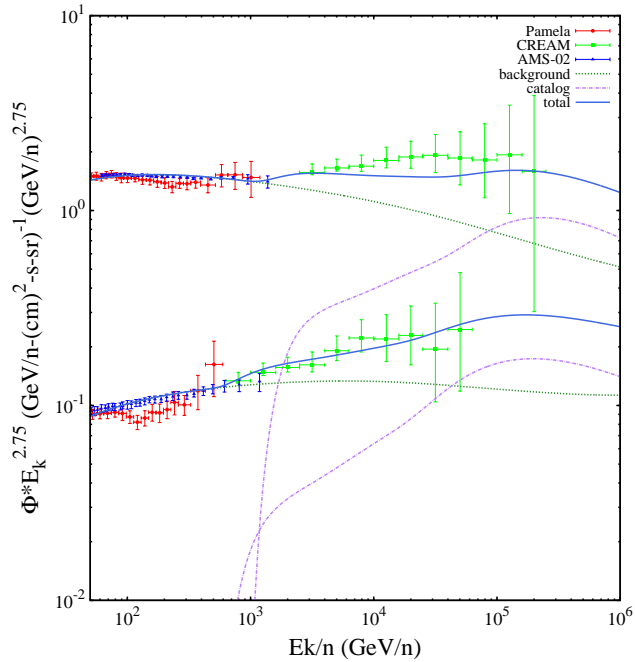}
\caption{
The best-fit model D is obtained by setting the explosion rate $\nu$ equal to 1.7 per century
and by letting the escape index $\gamma$ vary between 2 and 3. The latter adjusts itself to
a value of 2.36 close to 2, hence a kink in the proton and helium spectra at $\sim$ 1~TeV/nuc.
}
\label{fig:fit_D}
\end{figure}

\begin{figure}
\centering
\includegraphics[width=0.6\textwidth]{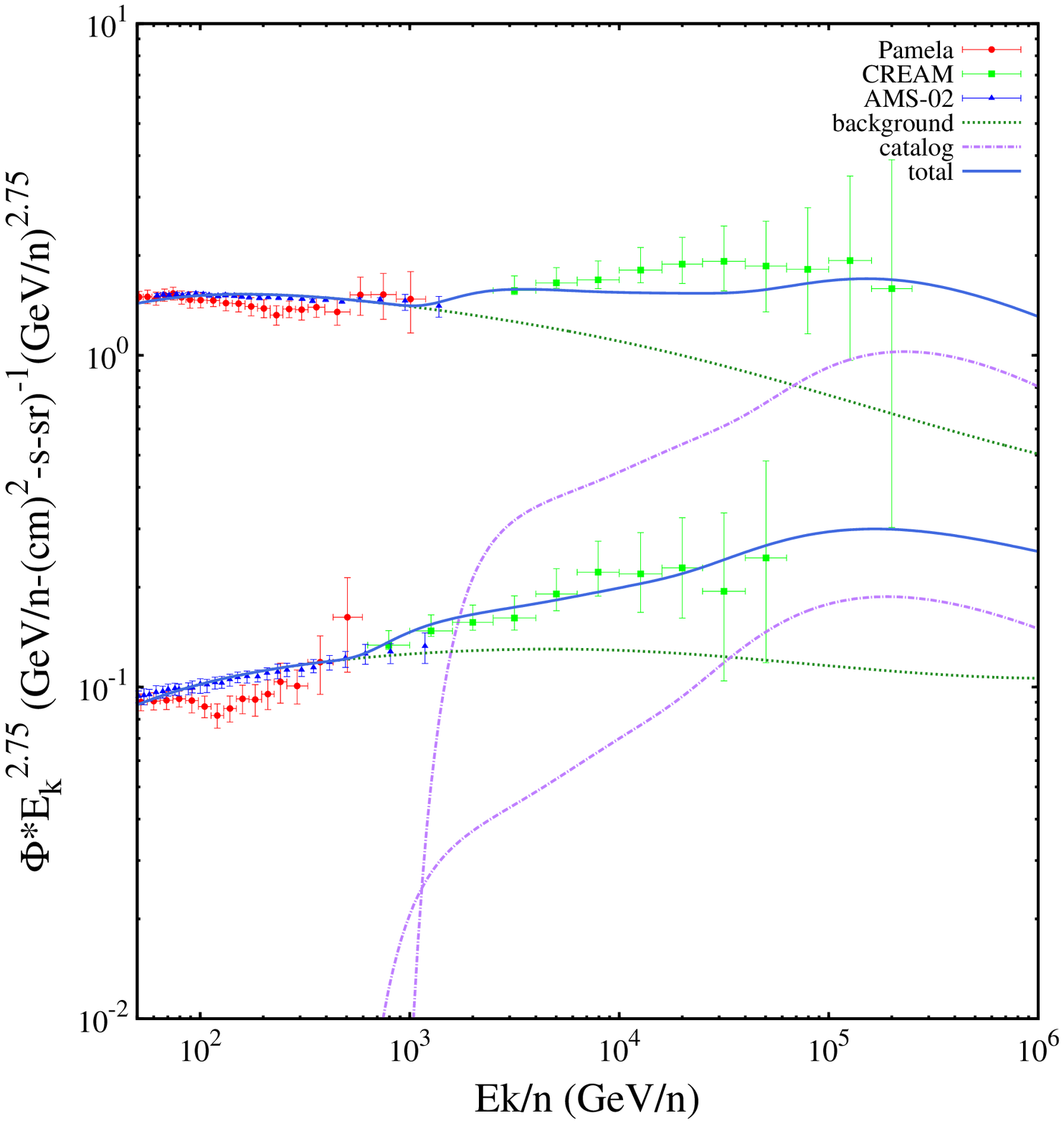}
\caption{
The previous fit is improved in model E where the explosion rate $\nu$ is now decreased
to a value of 1.5 per century.
}
\label{fig:fit_E}
\end{figure}

Finally, we consider the injection index $\gamma$ as a free parameter
and let it vary between $2$ and $3$. Model D (Fig.~\ref{fig:fit_D}) and
Model E (Fig.\ref{fig:fit_E}) is for SN explosion rates of 1.7 and
1.5 per century, respectively.
These values are very close to the average determined by \citet{2006Natur.439...45D}.
Further increasing the explosion rate would make fitting worse, especially for the proton
spectrum.


\section{Conclusion} \label{con}

The spectral hardening exhibited by the cosmic ray proton and helium fluxes above
a few hundreds of GeV/nuc in the PAMELA\citep{2011Sci...332...69A} data is no
longer present in the AMS-02\citep{ams2} observations which point toward a
power-law behavior. But the CREAM\citep{2010ApJ...714L..89A, 2011ApJ...728..122Y}
experiment still reports an excess above $\sim$ 1~TeV/nuc which is hard to understand
in the conventional model of cosmic ray transport and the problem persists.
We propose here a solution in terms of the known local and young SNRs whose
contributions become dominant above TeV energies. These sources have been extracted
from astronomical catalogs. The cosmic ray transport model has also been set to the
MED configuration which best-fits the B/C ratio. The thickness of the diffusive
halo is 4~kpc, a value that has been so far considered as canonical.

In this paper, we have improved over our previous analysis in three respects.
To commence, we have used the AMS-02 data which are much more difficult
to accommodate with a TeV spectral hardening than the PAMELA observations. The
goodness of our fits suffers from the power-law behavior of the AMS-02 measurements
as well as from smaller error bars associated with a significant improvement in the accuracy
of the data.
In spite of this, we get reduced chi-square values which are still satisfactory. To do so,
we have introduced two major revisions for the sources of our model. Supernova
explosions are distributed along spiral arms as they should in a typical SBc galaxy as
the Milky Way. The previous axi-symmetric distribution of CR background sources has
thus been replaced by a spiral distribution.
Then, we have explicitly taken into account the finite size of the local sources, for which
this effect is the most severe, and we have also modelled the energy-dependent escape
of cosmic rays from them.

We find that considering a spiral distribution for the background CR sources only leads to
limited improvements. On the contrary, a better description of local sources induces obvious
effects. For the canonical MED model of CR transport parameters, the average explosion rate
$\nu$ can be as large as $1.5$ per century, 
or even $1.7$ for model D, and becomes close to the fiducial value
of $1.9 \pm 1.1$ found by \citet{2006Natur.439...45D}. This is a significant improvement over
the \citet{2013A&A...555A..48B} analysis where $\nu$ had to be as small as 0.8 per century,
with a reduced chi-square value of 1.3 based on the PAMELA data, to be compared to our
best-fit result of 1.47 obtained with the more constraining AMS-02 measurements.
We confirm that a higher explosion rate reduces the role of local sources as is clear in
Table \ref{ta:model} where the proton and helium yields $q^0_{\text{p}}$ and $q^0_{\text{He}}$
are anti-correlated with $\nu$.
Our best-fit models D and E feature a characteristic kink at a few TeV/nuc. Should this hardening
be confirmed by future observations, it would point toward sources where the release of cosmic
rays in interstellar space cannot be considered as instantaneous, the most energetic particles
being emitted first.

\normalem
\begin{acknowledgements}
This work
is supported by the Ministry of Science and Technology 863 project 2012AA121701,
by the Chinese Academy of Science Strategic Priority Research Program ``The Emergence of
Cosmological Structures'' of the Chinese Academy of Sciences, Grant No. XDB09000000,  and
the NSFC grant 11373030. This work has also been supported by Institut universitaire de France.
\end{acknowledgements}

\bibliographystyle{raa}
\bibliography{ref_cr}
\end{document}